\documentclass[12pt]{article}
\usepackage{a4wide,graphics,graphicx,amsmath,amssymb,cite,nicefrac}
\usepackage[verbose]{wrapfig}
\usepackage{authblk,hyperref}
\usepackage{url,amsfonts}
\usepackage[makeroom]{cancel}
\catcode`@=11 \@addtoreset{equation}{section} \catcode`@=12

\begin{document}

\begin{titlepage}
\begin{center}

\vskip 1.0cm

{\bf \huge Penrose limits in massive type-IIA  \\ \vskip 5pt AdS$_3$ background}

\vskip 2.0cm

{\bf \large Sourav Roychowdhury${}^{1}$ and Prasanta K. Tripathy${}^{2}$}

\vskip 30pt
 {\it ${}^1$Chennai Mathematical Institute, \\
SIPCOT IT Park, Siruseri 603 103, India}\\\vskip 5pt
{\it ${}^2$%
Department of Physics, Indian Institute of
Technology
Madras,  \\ Chennai
600 036, India}\\

\vskip 10pt

\texttt{%
souravroy@cmi.ac.in},
\texttt{%
prasanta@iitm.ac.in}

\end{%
center}

\vskip 1.5cm

\begin{%
center} {\bf ABSTRACT}\\[3ex]\end{center}%

In this paper we consider the non-Abelian T-dual geometry of the type
$IIB$ supergravity theory on $AdS_3\times S^3\times T^4$ background along 
a convenient $SU(2)$ subgroup of the $SO(4)$ R-symmetry. We examine various 
null geodesics of the resulting massive type $IIA$ supergravity theory and 
investigate the Penrose limits along these geodesics. We find that one of 
the resulting backgrounds admits pp-wave geometry in the neighbourhood of 
a suitable null geodesic. We carry out the supersymmetry analysis of the 
resulting pp-wave geometry and observe that it preserves sixteen supercharges. 
Further we comment on the possible gauge theory dual of the resulting
pp-wave background.




%
\vfill


\end{titlepage}

\newpage 

\section{\label{Intro}Introduction}

The gauge-gravity correspondence \cite{Maldacena:1997re,Witten:1998qj} plays an extremely important role in
relating seemingly different theories each other. The BMN correspondence \cite{Berenstein:2002jq}
provides one such example where the string theory on a pp-wave background is 
related to a large $R$ charge sector of the ${\cal N}=4$ superconformal
theory in four dimensions. The BMN correspondence is used to compute 
interacting string states from perturbative gauge theory \cite{Berenstein:2002jq,Sadri:2003pr}. The pp-wave background results upon taking the Penrose limit of supergravity theory in ten and eleven dimensions 
\cite{blau.m,Figueroa-OFarrill:2001hal,Blau:2001ne,Blau:2002dy}. 
They provide an exact string theory background to all orders of $\alpha'$ 
and $g_s$ \cite{Amati:1988sa,Horowitz:1989bv}. The exact nature of the pp-wave background enables one to compute 
the interacting string states using the BMN correspondence \cite{Berenstein:2002jq}. In view of this, 
exploring pp-wave geometries as the Penrose limits of various supergravity 
theories in ten and eleven dimensions have become an important arena of 
research during the past several decades.

A particularly interesting set of backgrounds which are of recent 
interest in this context are the ones obtained from the conventional
supergravity theories by applying non-Abelian T-duality \cite{delaOssa:1992vci}. Non-Abelian
T-duality is a generalization of the standard Abelian T-duality, where 
a non-Abelian isometry group of a supergravity solution is used for dualization.
However, unlike the Abelian case, it is not a symmetry of the full 
string theory. It is mainly used as a solution generating technique 
to obtain new supergravity backgrounds from know ones. Initially 
non-Abelian T-duality was formulated for supergravity backgrounds with 
NS-NS fields. However, this duality has received a wide range of 
applicability after it has been generalized to incorporate the RR 
sector \cite{Sfetsos:2010uq}. Non-Abelian T-dual backgrounds has been constructed for a 
large class of theories and their relevance in the gauge-gravity
correspondence has been envisaged \cite{Lozano:2016kum,Lozano:2016wrs,Itsios:2012zv,Itsios:2013wd,Araujo:2015npa,Whiting,Itsios:2017cew,Gaillard:2013vsa,Lozano:2019ywa,Lozano:2015bra,Itsios:2012dc,Lozano:2019zvg,Macpherson:2014eza}. The Penrose limits for some of 
these dual geometries has been analysed \cite{Itsios:2017nou,Roychowdhury:2019kqr,Roychowdhury:2021lji,Nastase:2021qvv}. It has been shown that 
a number of these geometries indeed admit pp-wave solutions upon
taking the Penrose limit. Of particular interest among these 
backgrounds are the non-Abelian duals of $AdS_5\times S^5$
background \cite{Itsios:2017nou}, the Klebanov-Witten background \cite{Roychowdhury:2019kqr} and the Klebanov-Tseytlin
background \cite{Roychowdhury:2021lji}. Some of these dual background admit pp-wave geometries
even when the original geometry before dualization provide no such 
solution \cite{PandoZayas:2002dso,Roychowdhury:2021lji}.

Most of the above results focus on pp-waves geometries in non-Abelian 
T-dual geometries containing an $AdS_5$ factors. However there has not
been much progress in analysing the Penrose limits of backgrounds 
containing an $AdS_3$ factor. In the present work we plan to consider
one such background and inspect the Penrose limits in it. We focus 
on the non-Abelian T-dual of type $IIB$ supergravity theory on 
$AdS_3\times S^3\times CY_2$. This background has been constructed
recently and its gauge theory dual has also been studied \cite{Lozano:2019ywa}. Here we 
show that  this geometry indeed admits pp-wave solution upon taking
the Penrose limit in the vicinity of appropriate null geodesics.
The plan of the paper is as follows. In the following section, we first 
briefly review the background. Subsequently we consider the Penrose
limits along appropriate null geodesics and obtain the pp-wave geometry.
In \S3 we analyse the spinor conditions and show that the resulting 
pp-wave solution preserves sixteen supercharges. Finally in \S4 we comment
upon the possible gauge theory dual before concluding our results 
in \S5. Some technical details are outlined in the appendix.

\section{The non-Abelian T-dual of AdS$_3 \times$ CY$_2 \times$ S$^3$ and its Penrose limits}

In this section we will first focus on the non-Abelian T-dual of the type $IIB$
supergravity on the background of $AdS_3 \times CY_2 \times S^3$. This geometry 
arises as the near horizon limit of a stack of intersecting $D1-D5$ configuration.
This is one of the simplest examples used to apply non-Abelian T-duality in
the presence of background RR-fields. The metric is given by
\begin{eqnarray}  \label{main metric}
ds^2 = 4L^2 ds^2 \big(AdS_3\big) + L^2 ds^2 \big(CY_2\big) + 4L^2 ds^2 \big(S^3\big) \ .
\end{eqnarray}
where
\begin{eqnarray} \label{ads3 s3} 
&ds^2 \big(AdS_3\big) =& -\cosh^2r dt^2 + dr^2 + \sinh^2r  \ d\chi^2  \ , \nonumber\\
& ds^2 \big(S^3\big)=&  d\alpha^2 + d\beta^2 + d\gamma^2 + 2 \cos\alpha d\beta d\gamma \ .
\end{eqnarray}
and $ds^2(CY_2)$ represents the metric of a Calabi-Yau twofold. Here the parameter $L$ represents the size 
of the internal manifold $CY_2$. In the present work we will entirely focus 
on the internal manifold $T^4$.  We denote
$z_i,\ i = 1,2,3,4 $ to parameterize the $T^4$. Thus,
\begin{eqnarray} \label{T4}
ds^2 \big(T^4\big) = \sum_{i=1}^4 dz_i^2 \ . 
\end{eqnarray}
In addition to the above, the supergravity background is supported by the 
dilaton field $e^{2\Phi} = 1$  and the RR three-form flux 
\begin{eqnarray} \label{main fields}
 F_3 =  8L^2 \ \text{Vol} \big(S^3\big) \ . 
\end{eqnarray}

We will now consider the non-Abelian T-dual of the above background \cite{Sfetsos:2010uq}. The 
background \eqref{main metric}-\eqref{main fields} has an $SO(4)$ 
R-symmetry realised by the presence of $S^3$ factor in the background metric. The authors of \cite{Sfetsos:2010uq} considered an $SU(2)$ 
subgroup of this R-symmetry group to perform the duality transformation.
The T-dual background is specified by the metric:
\begin{eqnarray} \label{natdm}
d\hat{s}^2_{\text{NATD}} = 4L^2 ds^2 \big(AdS_3\big) + L^2 ds^2 \big(T^4\big) + \frac{L^2}{4} d\rho^2 + \frac{L^2 \rho^2}{4 + \rho^2} \ d\Omega_2^2\big(\theta, \phi) \ ,
\end{eqnarray}
where $d\Omega_2^2\big(\theta, \phi) = d\theta^2 + \sin^2\theta d\phi^2$ is 
the round metric on the 2-sphere. Here we have set $\alpha^{\prime} =1$ for
convenience. Also we have rescaled the $\rho$-coordinate as $\rho \rightarrow 
L^2 \rho$ in order to get an overall $L^2$ factor in the metric.

The remaining fields in the $NS-NS$ sector  of the dual background  are given 
by the expressions 
\begin{eqnarray} \label{natd ns} 
e^{-2\hat{\Phi}} = \frac{L^6}{4 g_{s}^2} \big(4 + \rho^2\big) \ ,  \  \hat{B}_2 = - \frac{L^2 \rho^3}{2\big(4 + \rho^2\big)} \sin\theta  d\theta \wedge d\phi \   ,
\end{eqnarray}
whereas the RR sector is described by the field strengths
\begin{eqnarray} \label{natdrr} 
\hat {F}_0 = \frac{L^2}{g_s} \ , \ \hat{F}_2 = - \frac{L^4 \rho^3}{2 g_s \big(4 + \rho^2\big)} \sin\theta  d\theta \wedge d\phi \  ,  \  \hat{F}_4 = - \frac{L^6}{g_s} \text{Vol} \big(T^4\big) \ .
\end{eqnarray}

This geometry corresponds to a massive type $IIA$ supergravity on 
$AdS_3\times T^4\times M_3$. It has a $SU(2)$ global symmetry and 
it preserves only half the supersymmetry of the original background.
The manifold $M_3$ consists of a $S^2$ fibration over a half line.
For small $\rho$ it resembles $\mathbb{R}^3$ where as for large 
$\rho$ it has the product form $S^2\times\mathbb{R}$. It calls 
for recent interest in the context of two dimensional defect CFTs
where this geometry appears as the leading order in an expansion
of the number of gauge groups of the dual CFT \cite{Lozano:2019ywa}.

We will now turn our discussion on the Penrose limits. For the 
$AdS_3 \times S^3 \times T^4$ background the Penrose limits have been 
extensively studied in \cite{Cho:2002zp}. The authors showed that the
Penrose limits indeed give rise to pp-wave geometry. They have also 
considered  Abelian T-duality of the background geometry along a 
Hopf-fibre of $S^3$ and examined the Penrose limits in the resulting
Abelian T-dual background \cite{Cho:2002zp}. 

Here we will examine the Penrose limits for the non-Abelian T-dual
background. We will consider appropriate null geodesics and take 
the Penrose limits along these geodesics. Consider first the geodesic
equation
\begin{eqnarray} \label{geo}
 \frac{d^2 x^\mu}{du^2} + \Gamma_{\nu \rho}^\mu  \frac{dx^\nu}{du} \frac{dx^\rho}{du} = 0 \ .
\end{eqnarray}
Here $\{x^\mu\}$ denote the spacetime coordinates and $u$ is the affine 
parameter along the geodesic. Let $x^\lambda$ denotes an isometry direction.
Then, for motion along this isometry direction, the geodesic equation 
reduces to the following simple form
\begin{eqnarray} \label{geo-2} 
\partial^\mu g_{\lambda \lambda} = 0 \ .
\end{eqnarray}
This is due to the fact that for motion along such an isometry direction the 
velocity as well as the acceleration along any $x^\mu$, $\mu \neq \lambda$ 
will be zero:
\begin{eqnarray} \label{geo-1} 
 \frac{dx^\mu}{du} = 0 = \frac{d^2 x^\mu}{du^2} \ ,  \ \text{for} \ \mu \neq \lambda  \ .
\end{eqnarray}
In order to obtain the Penrose limit we need in addition to focus in the 
vicinity of null geodesics. Thus, we impose $ds^2 = 0$ in addition to 
the geodesic condition \eqref{geo-2}.

We will now focus on various isometry directions of the background of interest
\eqref{natdm}. Let us first consider the $z_i$'s along the $T^4$. Though, these
form isometries of the metric, the geodesic condition for them is trivially
satisfied. Moreover we can write the $T^4$ as $\mathbb R\times S^3$ and consider
isometries of this $S^3$.  However, the geodesic condition for any of these isometries 
do not give any 
non-trivial constraint. Thus we will no longer consider these directions any further. This
leaves behind the motion along the $\phi$-direction which we will now focus.
The geodesic equation for this case becomes
\begin{eqnarray} \label{geo-3}
\partial_\mu g_{\phi\phi} = 0 \ .
\end{eqnarray}
This on the other hand gives nontrivial solutions as we can notice from the
relevant metric component $g_{\phi\phi}$ in \eqref{natdm}:
\begin{eqnarray} \label{g phi phi} 
 g_{\phi\phi} = \frac{L^2 \rho^2}{4 + \rho^2} \sin^2\theta  \ .
\end{eqnarray}
This has a nontrivial dependence on $\rho$ and $\theta$. 
Now for $\mu = \rho$,  the geodesic condition \eqref{geo-3} gives $\rho = 0$ and $\theta = \{0, \pi\}$.
Similarly setting $\mu = \theta$, we obtain $\rho = 0$ and $\theta = \{0, \frac{\pi}{2}, \pi\}$. However, $\rho = 0$ and $\theta = \{0, \pi\}$ lead to singular geometries 
as $g_{\phi\phi}$ component vanishes for all these geodesics.
Thus, we will not  consider the Penrose limits for such singular geometries. 

This leaves us the only possibility of considering the motion of a particle
carrying nonzero angular momentum in the $(\rho, \phi)$ plane. We will 
consider null geodesics for such motion in the neighbourhood of $r = 0 = z_i$ 
and $\theta = \frac{\pi}{2}$. 
The Lagrangian for a massless particle with background metric $g_{\mu\nu}$ is
\begin{eqnarray} \label{lag} 
\mathcal{L} = \frac{1}{2} g_{\mu\nu} \dot{X}^\mu \dot{X}^\nu  \ .
\end{eqnarray}
In the above we consider the affine parameter $u$ and the dots denote 
derivative with respect to it. We consider the background \eqref{natdm}.
Substituting the explicit expression for the metric we find 
\begin{eqnarray} \label{explicit lag}
\mathcal{L} = \frac{L^2}{2}  \bigg(-4 \dot{t}^2 + \frac{1}{4} \dot{\rho}^2 + \frac{\rho^2}{4 + \rho^2} \dot{\phi}^2\bigg) \ .
\end{eqnarray}

The above system is completely integrable and admits a one parameter 
family of solutions as we can see in the following.
Let us first analyse the symmetries of the above 
Lagrangian. Clearly, the coordinates $t$ and $\phi$ are cyclic. Hence 
the generalized momenta conjugate the to coordinates $t$ and $\phi$ 
are conserved. Consider the momentum conjugate to $t$ first:
\begin{eqnarray} \label{t}
\frac{\partial \mathcal{L}}{\partial \dot{t}} = - 4L^2 \dot{t} \ . 
\end{eqnarray}
Thus, we find $\dot t= {\rm const}$. Choosing the affine parameter $u$ 
suitably we set $\dot t = 1$. Now consider the equation of motion
pertaining to the $\phi$ coordinate:
\begin{eqnarray} 
\frac{\partial \mathcal{L}}{\partial \dot{\phi}} = L^2 \frac{\rho^2}{4 + \rho^2} \dot{\phi} = {\rm const} \ .
\end{eqnarray}
We introduce the constant $J$ to denote this conserved quantity:
\begin{equation} \label{J} 
J = - \frac{\rho^2}{4 + \rho^2} \dot{\phi} \ . 
\end{equation}
Finally, we need to consider the equation for $\rho$-coordinate. Requiring
the geodesics to be null {\it i.e.} $\mathcal{L} = 0$, we obtain
\begin{eqnarray} \label{rho} 
\dot{\rho}^2 = 4 \bigg(4 - \frac{4+\rho^2}{\rho^2} J^2\bigg)  \ .
\end{eqnarray}
This equation admits exact analytical solution which has a simple form:
\begin{eqnarray} \label{sol} 
\rho^2 = \frac{4J^2 + 4 \big(J^2 - 4\big)^2 \big(u + c_{\rho}\big)^2}{4 - J^2} \ ,   \  
\end{eqnarray}
with $c_\rho$ being the constant of integration. We can set it to zero by
redefining the affine parameter $u$ by a constant shift. Substituting the 
above in \eqref{J} and integrating we find 
\begin{eqnarray} 
\phi = \frac{1}{J} \Bigg[u - \tan^{-1} \bigg(\frac{(J^2-4)u}{J}\bigg)\Bigg]   \ ,   
\end{eqnarray}

We will now obtain the Penrose limit for the above null geodesic carrying 
an angular momentum $J$ around  $r = 0 = z_i$ and $\theta = \frac{\pi}{2}$.
First, we redefine the coordinates as follows 
\begin{eqnarray} \label{redefine}
r = \frac{\bar{r}}{L} \ , \ z_i = \frac{y_i}{L} \ ;  \ i =1,2,3,4 \ ,  \     \ \theta =  \frac{\pi}{2} + \frac{x}{L} \ . 
\end{eqnarray}
In addition to the above, we rescale the string coupling as $g_s = L^3 \tilde{g}_s$, in order to keep the dilaton finite upon taking the Penrose 
limit. Finally, we will consider the following large $L$ expansion:
\begin{eqnarray} \label{expand}
dt = c_1 du \ , \ d\phi = c_2 du + c_3 \frac{dw}{L} + c_4 \frac{dv}{L^2} \ , \ d\rho = c_5 du + c_6 \frac{dw}{L} \ . 
\end{eqnarray}
The expansion includes the coefficients $c_i$ which we need to determine. 
Notice that the geodesic to be null determines three of the coefficients 
as follows:
\begin{eqnarray} \label{c}
c_1 = 1 \ ,  \ c_2 = - \frac{4+\rho^2}{\rho^2} J \ , \ c_5 = 2  \Bigg[4 - \frac{\big(4+\rho^2\big)}{\rho^2} J^2\Bigg]^{\frac{1}{2}} \ .
\end{eqnarray}

We will now substitute the expansion \eqref{expand} in the T-dual metric
\eqref{natdm} and take the limit $L\rightarrow\infty$. It may be noted 
that the expansion contains terms of order $\mathcal{O}(L^2)$. However,
they cancel each other upon imposing the null geodesic condition. In 
addition, it also contains terms of order $\mathcal{O}(L)$. Requiring 
the metric to be finite imposes the following condition on the 
coefficients $c_i$:
\begin{eqnarray}
c_5 c_6 + \frac{4\rho^2}{4 + \rho^2} \ c_2 c_3 = 0 \ . 
\end{eqnarray}
We can substitute the values of $c_2$ and $c_5$ from \eqref{c} in the above equation. This leads to the following relation between the coefficients 
$c_3$ and $c_6$:
\begin{eqnarray}
\frac{c_3}{c_6} = \frac{1}{2J} \Bigg[4 - \frac{\big(4+\rho^2\big)}{\rho^2} J^2\Bigg]^{\frac{1}{2}} \ . 
\end{eqnarray}
Subsequently we will show that upon requiring the background to satisfy the
Einstein's equations determines the coefficient $c_3$. This leaves behind
the only undetermined coefficient $c_4$. Choice of an appropriate normalization
for the cross term $du dv$ in the metric determines the value of this coefficient to be $c_4 = - \frac{1}{J}$. 

We will now substitute the above results in the background metric \eqref{natdm} and take the limit $L \rightarrow \infty$. The leading term of the resulting
metric takes the form
\begin{eqnarray} \label{met}
&ds^2 =& 2 dudv + 4 d\bar{r}^2 + 4 \bar{r}^2 d\chi^2 + dy_1^2 + dy_2^2 + dy_3^2 + dy_4^2  + \bigg(\frac{c_6^2}{4} + \frac{\rho^2}{4+\rho^2} c_3^2\bigg) dw^2 \nonumber\\
&& \ +  \ \frac{\rho^2}{4+\rho^2} dx^2 - \bigg(4 \bar{r}^2  + \frac{4+\rho^2}{\rho^2} J^2 x^2\bigg) du^2 \ .
\end{eqnarray}
This solution indeed corresponds to a pp-wave geometry. However, it is not in
the standard Brinkmann form. Subsequently we will rewrite it in the Brinkmann
form upon suitable redifinition of the coordinates. Before that, we will 
consider the Penrose limit for the remaining background fields. It is 
straightforward to show that the NS-NS two-form $\hat B_2$ and the dilaton 
$\hat\Phi$ takes the form
\begin{eqnarray}
&\hat{B}_2 =& \frac{c_3 \rho^3}{2\big(4+\rho^2\big)}   \ dw \wedge dx  \ ,  \nonumber\\  
&e^{-2\hat{\Phi}} =& \frac{1}{4 \tilde{g}_{s}^2} \big(4 + \rho^2\big) \ .  
\end{eqnarray}
In obtaining the expression for $\hat B_2$ we have ignored exact terms which
can be gauged away. The field strength corresponding to the NS-NS two-form 
$\hat B_2$ is
given by 
\begin{eqnarray}
\hat{H}_3 =  \rho \  \frac{\sqrt{4\rho^2 - \big(4 + \rho^2\big) J^2}}{\big(4 + \rho^2\big)^2 }  \ \Bigg[\big(12 + \rho^2\big) c_3 + \rho \big(4 + \rho^2\big) c_3^{\prime} \Bigg] \ du \wedge dw \wedge dx \ .    
\end{eqnarray}
In the above we have used the result $d\rho = c_5 du$ where the expression
for the coefficient $c_5$ is given by \eqref{c}. We can similarly carry out
the Penrose limit for the fields corresponding to the RR sector. We find 
that the field strengths $\hat F_0$ and $\hat F_4$ vanish whereas the 
RR two-form field strength $\hat F_2$ takes the simple form
\begin{eqnarray}
\hat{F}_2 =  \frac{J \rho}{2 \tilde{g}_s}   \ dx \wedge du \ .
\end{eqnarray}

We will now transform the metric \eqref{met} into the standard Brinkmann
form \cite{blau.m} using the formalism developed in \cite{Itsios:2017nou}.
We will first consider a metric of the form
\begin{eqnarray} \label{ele}
ds^2 = 2 dudv \ + \sum_i A_i (u) \ dx_i^2 \ .
\end{eqnarray}
Notice that the resulting pp-wave background \eqref{met} is indeed of the 
above form for suitable choice of the functions $A_i(u)$. We will now 
redefine the transverse coordinates $x_i$ as 
\begin{eqnarray}
x_i \rightarrow \frac{x_i}{\sqrt{A_i}} \ .    
\end{eqnarray}
Further we make the following redefinition of one of the lightcone coordinates 
\begin{equation}
v \rightarrow v + \frac{1}{4} \sum_i \frac{\dot{A}_i}{A_i} \ x_i^2 \ .
\end{equation}
The resulting metric is in the Brinkmann form
\begin{eqnarray}
ds^2 = 2 dudv \ + \sum_i dx_i^2  + \bigg(\sum_i F_i (u) \ x_i^2\bigg) du^2 \ ,  
\end{eqnarray}
with
\begin{eqnarray} \label{F} 
F_i =  \frac{1}{4} \frac{\dot{A}_i^2}{A_i^2} + \frac{1}{2} \frac{d}{du} \bigg(\frac{\dot{A}_i}{A_i}\bigg) \ .
\end{eqnarray}

For the present case we can read the functions $A_i(u)$ from our 
pp-wave metric \eqref{met} as 
\begin{eqnarray}
A_{\bar{r}} = 4 \ , \ A_w = \frac{c_6^2}{4} + \frac{\rho^2}{4+\rho^2} c_3^2 \ , \  A_x = \frac{\rho^2}{4+\rho^2} \ .
\end{eqnarray}
Now we need to make the redefinition
\begin{eqnarray}
&&\bar{r} \rightarrow \frac{\bar{r}}{\sqrt{A_{\bar{r}}}} \ ,  \ w \rightarrow \frac{w}{\sqrt{A_w}} \ ,  \ x \rightarrow \frac{x}{\sqrt{A_x}} \ \ \text{and}               \nonumber\\
&& v \rightarrow v + \frac{1}{4} \Bigg[\frac{\dot{A}_w}{A_w} \ w^2  + \frac{\dot{A}_x}{A_x} \ x^2\Bigg]  \ , 
\end{eqnarray}
to obtain
\begin{eqnarray} \label{metric brinkmann}
&ds^2 =& 2 dudv + d\bar{r}^2 + \bar{r}^2 d\chi^2 + dy_1^2 + dy_2^2 + dy_3^2 + dy_4^2 +  dw^2 + dx^2 \nonumber\\
&&- \ \Bigg[\bar{r}^2  + \bigg(\frac{(4+\rho^2)^2}{\rho^4} J^2 - F_x\bigg) x^2 - F_w w^2\Bigg] du^2 \ .
\end{eqnarray}
The functions $F_i$ are determined using the expression \eqref{F}.

In the Brinkmann coordinates the NS-NS fields take the form
\begin{eqnarray} \label{NS-NS Brinkmann}
&e^{-2\hat{\Phi}}=& \frac{1}{4 \tilde{g}_{s}^2} \big(4 + \rho^2\big) \ , \nonumber\\
&\hat{H}_3 =&  \frac{\sqrt{4\rho^2 - \big(4 + \rho^2\big) J^2}}{\big(4 + \rho^2\big)^{\frac{3}{2}} }  \ \Bigg[\big(12 + \rho^2\big) c_3 + \rho \big(4 + \rho^2\big) c_3^{\prime} \Bigg]   \Bigg[\frac{c_6^2}{4} + \frac{\rho^2}{4+\rho^2} c_3^2\Bigg]^{- \frac{1}{2}}    \nonumber\\
 &&du \wedge dw \wedge dx \ .
\end{eqnarray}
Similarly, for the RR sector, we find
\begin{eqnarray} \label{Brinkmann RR}
 &&\hat{F}_2 =  \frac{J}{2\tilde{g}_s}   \sqrt{4 + \rho^2} \ dx \wedge du \ , \nonumber\\
 && \hat{F}_0  = 0 = \hat{F}_4 \ . 
\end{eqnarray}

We will now verify and show that the above background indeed satisfies the
equations of motion. We will first consider the Bianchi identities and the
gauge field equations. Subsequently we will turn our attention to the 
Einstein's equations. A quick inspection of the background fields 
\eqref{NS-NS Brinkmann}-\eqref{Brinkmann RR} shows that the field 
strengths $\hat{H}_3$ and $\hat F_2$ are indeed closed. In addition 
we have $\hat{F}_0  = 0 = \hat{F}_4 $. Thus the 
Bianchi identities
\begin{eqnarray} \label{Bianchi 2a natd}
d\hat{H}_3 = 0 \ , \  d\hat{F}_2 = \hat{F}_0 \hat{H}_3 \ , \ d\hat{F}_4 = \hat{H}_3 \wedge \hat{F}_2 \ 
\end{eqnarray}
are trivially satisfied.

We will now turn our attention to the type-$IIA$ supergravity equations:
\begin{eqnarray} \label{Bianchi 2a natd1}
&&d\Big(e^{-2\hat{\Phi}} \star \hat{H}_3\Big) - \hat{F}_2 \wedge \star \hat{F}_4 - \frac{1}{2}  \hat{F}_4 \wedge \hat{F}_4 = \hat{F}_0 \star \hat{F}_2 \ , \nonumber\\
&&d\star \hat{F}_2 + \hat{H}_3 \wedge \star \hat{F}_4 = 0 \ , \nonumber\\
&&d\star \hat{F}_4 + \hat{H}_3 \wedge \hat{F}_4 = 0 \ .
\end{eqnarray}
For the background fields \eqref{NS-NS Brinkmann}-\eqref{Brinkmann RR}, the
above set of equations reduces to 
\begin{eqnarray}
&&d\Big(e^{-2\hat{\Phi}} \star \hat{H}_3\Big) = 0 \cr
&& d\star \hat{F}_2 = 0 \ .
\end{eqnarray}
To verify that they hold, consider the Hodge duals of the field strengths
$\hat H_3$ and $\hat F_2$:
\begin{eqnarray} \label{Hodge dual}
&\star \ \hat{H}_3=& \frac{\sqrt{4\rho^2 - \big(4 + \rho^2\big) J^2}}{ \big(4 + \rho^2\big)^{\frac{3}{2}} }  \ \Bigg[\big(12 + \rho^2\big)c_3 + \rho \big(4 + \rho^2\big) c_3^{\prime} \Bigg]   \Bigg[\frac{c_6^2}{4} + \frac{\rho^2}{4+\rho^2} c_3^2\Bigg]^{- \frac{1}{2}} \nonumber\\
&& du \wedge d\Omega_2 \wedge dy_1 \wedge dy_2 \wedge dy_3 \wedge dy_4 \ , \nonumber\\
&\star \ \hat{F}_2=& \frac{J}{2\tilde{g}_s}   \sqrt{4 + \rho^2}  \ du \wedge d\Omega_2 \wedge dy_1 \wedge dy_2 \wedge dy_3 \wedge dy_4 \wedge dw \ . 
\end{eqnarray}
Using the above expression it can be shown $\star\hat H_3 \ , \Big(e^{-2\hat\Phi}\star\hat H_3\Big)$ as well as $\star \ \hat{F}_2$ are all closed. This is 
due to the fact that all the nonvanishing components of the Hodge dual are 
functions of the $\rho$-coordinate only. Because of the presence of $du$
factor both the Hodge duals are all closed forms. It is interesting to see
that the gauge field equations are satisfied irrespective of the value 
of the undetermined coefficient $c_3$. However this is not the case for 
the Einstein's equations as we will see in the following.

Consider the Einstein's equations for type $IIA$ supergravity:
\begin{eqnarray} \label{Einstein eq natd}
\hat{R}_{\mu\nu} + 2D_{\mu}D_{\nu}\hat{\Phi} = \frac{1}{4} \hat{H}_{\mu\nu}^2 + e^{2\hat{\Phi}} \Bigg[\frac{1}{2} (\hat{F}_2^2)_{\mu\nu} + \frac{1}{12} (\hat{F}_4^2)_{\mu\nu}- \frac{1}{4} g_{\mu\nu} \Big(\hat{F}_{0}^2 + \frac{1}{2} \hat{F}_{2}^2 + \frac{1}{4!}\hat{F}_{4}^2 \Big)\Bigg]  \ , 
\end{eqnarray}
along with the equation of motion for the dilaton
\begin{eqnarray} \label{R eq natd}
\hat{R} + 4D^2\hat{\Phi} - 4(\partial\hat{\Phi})^2 - \frac{1}{12}\hat{H}^2=0 \ .
\end{eqnarray}
A careful analysis of these equations show that the equation of motion for
the dilaton is automatically satisfied for our pp-wave background. In addition the Einstein's equations
automatically hold for all the values of $\mu,\nu$ except for the case $\mu=\nu=u$, where
the equation takes form
\begin{eqnarray} \label{Einstein uu}
\hat{R}_{uu} + 2D_{u}D_{u}\hat{\Phi} = \frac{1}{4} \hat{H}_{uu}^2 + \frac{1}{2} e^{2\hat{\Phi}}  \Big(\hat{F}_2^2\Big)_{uu}   \ .
\end{eqnarray}
This indeed provides a nontrivial constraint involving the coefficient $c_3$.
The coefficient $c_3$ can be determined upon solving this equation. Thus, 
the Einstein's equation determines the coefficient and the pp-wave background
satisfies the equations of motion. In the appendix we outline the computation
involving the Einstein's equation leading to the determination of the 
coefficient $c_3$.

\section{Supersymmetry of pp-wave} 

In this section we will consider the supersymmetry preserved by our pp-wave
background. In the case of backgrounds with an $AdS_5$ factor such as 
$AdS_5\times S^5$, non-Abelian T-duality breaks the supersymmetry from
$\mathcal{N}=4$ to $\mathcal{N}=2$ \cite{Lozano:2016kum}. Whereas, for the case of Klebanov-Witten as well as the Klebanov-Tseytlin background, the dual background has $\mathcal{N}=1$ supersymmetry \cite{Itsios:2012zv,Itsios:2013wd}. Because, for the Klebanov-Witten and Klebanov-Tseytlin case the Killing spinor of the original background does not carry any $SU(2)$-charge that used for dualization \cite{Lozano:2011kb,Kelekci:2014ima}. 
However, in some of these examples it has been shown that the resulting 
pp-wave background restores some of the supersymmetries. For the present 
case the non-Abelian T-dual of $AdS_3\times S^3\times T^4$ itself preserves 
half the supersymmetries of the original background \cite{Sfetsos:2010uq}. 
Our goal here is to analyse the spinor conditions in order to obtain the 
number of supersymmetries preserved by the pp-wave background as given in 
\eqref{metric brinkmann}-\eqref{Brinkmann RR}.

To analyse the supersymmetry conditions, let us introduce  the Brinkmann 
coordinates $X^i$ such that
\begin{eqnarray} \label{x}
&& d\bar{r}^2 + \bar{r}^2  d\chi^2 = \big(dX^i\big)^2 \ ; \ i =1,2 \ ,    \  y_i = X^i  \ ; \ i =3,4,5,6  \ , \nonumber\\
&&w = X^7 \ ,  \ x = X^8 \ .  \nonumber\\
\end{eqnarray}
With this notation, the pp-wave background \eqref{metric brinkmann}-\eqref{Brinkmann RR} takes the form 
\begin{eqnarray} \label{in X coordinates} 
&ds^2 =& 2 dudv + \sum_{i=1}^{8} dX_i^2 + \mathcal{H} \ du^2 \ ,  \nonumber\\
&\hat{\Phi}=& \hat{\Phi}(u) \ , \nonumber\\
&\hat{H}_3 =& f_1(u) \ du \wedge dX^7 \wedge dX^8 \ ,                       \nonumber\\
&\hat{F}_2 =& f_2(u) \ dX^8 \wedge du \ .                      
\end{eqnarray}
where, for easy reading, we have introduced the notations
\begin{eqnarray} \label{Hf} 
&\hat{\Phi}(u) =& \frac{1}{2} \ln \Bigg[\frac{4 \tilde{g}_s ^2}{4 + \rho^2}\Bigg] \ , \nonumber\\
&\mathcal{H} =& \sum_{i,j =1}^8  F_{ij} X^i X^j \ ;  \ F_{ij} = F_{ji} \ ,  \nonumber\\
&f_1(u) = & \frac{\sqrt{4\rho^2 - \big(4 + \rho^2\big) J^2}}{ \big(4 + \rho^2\big)^{\frac{3}{2}} }  \ \Bigg[\big(12 + \rho^2\big)c_3 + \rho \big(4 + \rho^2\big) c_3^{\prime} \Bigg]   \Bigg[\frac{c_6^2}{4} + \frac{\rho^2}{4+\rho^2} c_3^2\Bigg]^{- \frac{1}{2}}  \ ,                  \nonumber\\
&f_2(u) =& \frac{J}{2\tilde{g}_s}   \sqrt{4 + \rho^2}  \  . 
\end{eqnarray}
In the above, the functions $F_{ij}$ are given by the expressions
 \begin{eqnarray} \label{Fij}
 &&F_{11} = F_{22} = -1 \ ,   \nonumber\\
 &&F_{77} = F_ {w} \ , \ F_{88} =  F_x -  \frac{(4+\rho^2)^2}{\rho^4} J^2 \ . 
\end{eqnarray}
Further, we introduce the frame $\{e^a\}$ such that
\begin{eqnarray} \label{frame}
e^- = du \ , \ e^+ =  dv + \frac{1}{2} \mathcal{H} du \ ,  \  e^i = dX^i \ . 
\end{eqnarray}
The pp-wave metric \eqref{in X coordinates} in this frame is given by
\begin{eqnarray} \label{bgmet}
ds^2 = 2e^+e^-  + \sum_{i=1}^8 \big(e^i\big)^2 = \eta_{ab} e^a e^b \ , 
\end{eqnarray}
with $\eta_{+-} = \eta_{-+} =1$ and $\eta_{ij} = \delta_{ij}$. The
remaining background fields in the frame takes the form
\begin{eqnarray} \label{in terms of frame}
&\hat{\Phi}=& \hat{\Phi}(u) \ ,  \nonumber\\
&\hat{H}_3 =& f_1(u) \ e^- \wedge e^7 \wedge e^8 \ ,                       \nonumber\\
&\hat{F}_2 =& f_2(u) \ e^8 \wedge e^- \ .                       
\end{eqnarray}
We also need to use the spin connections in the supersymmetry conditions. 
From \eqref{frame}-\eqref{bgmet} it is easy to compute the spin connections. The 
non-vanishing components are given by 
\begin{eqnarray} \label{spin}
\omega_{-i}  = \omega^{+i} = \frac{1}{2} \partial_{i} \mathcal{H} \ du \ . 
\end{eqnarray}


Let us now focus on the spinor conditions in detail. The supersymmetric
 variations for the dilatino and gravitino are given by
\begin{eqnarray}  \label{SUSY} 
&& \delta \hat{\lambda} = \frac{1}{2} \cancel{\partial} \hat{\Phi}  \hat{\epsilon} - \frac{1}{24} 
\cancel{\hat{H}} \sigma_3  \hat{\epsilon} \ + \frac{3}{16} e^{\hat{\Phi}}  \cancel{\hat{F}}_2 \left(i \sigma_2\right)  \hat{\epsilon} \ , \nonumber\\
&& \delta \hat{\psi}_{\mu} = D_{\mu}\hat{\epsilon} - \frac{1}{8} \hat{H}_{\mu\nu\rho} \Gamma^{\nu\rho} \sigma_3 \hat{\epsilon} + \frac{1}{16} e^{\hat{\Phi}} \cancel{\hat{F}}_2 \left(i \sigma_2\right)  \Gamma_{\mu}\hat{\epsilon} \ .
\end{eqnarray}
In the above we use the conventions of \cite{Itsios:2012dc,Itsios:2017nou}. 
In particular, we note that the action of the covariant derivative $D_\mu$
on the Killing spinors is given as $D_{\mu}\hat{\epsilon} = \partial_{\mu}\hat{\epsilon} + \frac{1}{4} \omega_{\mu}^{ab} \Gamma_{ab} \hat{\epsilon}$. 
In addition we use the notation $\cancel{\hat{F}}_n \equiv \hat{F}_{i_{1}...i_{n}} \Gamma^{i_{1}...i_{n}}$ and follow the standard convension
for $\sigma_i$ to denote the Pauli matrices. The Killing spinor 
$\hat{\epsilon}$ is given as 
\begin{eqnarray} \label{killing}
\hat{\epsilon} = \left( \begin{array}{ccc}
\hat{\epsilon}_+ \\
\hat{\epsilon}_-\\
\end{array} \right) \ ,
\end{eqnarray}
in terms of two Majorana-Weyl spinors $\hat{\epsilon}_{\pm}$.
It satisfies the relation $\Gamma_{11} \hat{\epsilon} = - \sigma_3 \hat{\epsilon}$ .
where we denote 
\begin{eqnarray}  \label{gamma} 
\Gamma^{\pm} = \frac{1}{\sqrt{2}} \Big(\Gamma^9 \pm \Gamma^0\Big) \ . 
\end{eqnarray}

We will now focus on the dilatino variation. Substituting the pp-wave
metric and the background fields as given by \eqref{in terms of frame} and simplifying
we find 
\begin{eqnarray}  \label{dilatino} 
\Gamma^- \Bigg[\dot{\hat{\Phi}} - \frac{1}{2} f_1(u) \Gamma^{78}  \sigma_3 - \frac{3e^{\hat{\Phi}}}{4}  \ f_2(u) \Gamma^8  \left(i \sigma_2\right)\Bigg] \hat{\epsilon} = 0 \ . \nonumber\\
\end{eqnarray}
The above equation is satisfied for $\Gamma^- \hat{\epsilon} = 0$. We will now 
proceed to solve the spinor conditions pertaining to the gravitino variation.
We need to set $\delta\hat\psi_\mu=0$ for $\mu=+,-,i$. First consider the
variation $\delta\hat\psi_+$. From the expression for the NS-NS three form
flux, we find $\hat H_{+\mu\nu}=0$ for all $\mu,\nu$. Using it along with 
the condition $\Gamma_{+} \hat{\epsilon} = \Gamma^{-} \hat{\epsilon} = 0$
in the variation $\delta\hat{\psi}_+=0$ we find that the Killing 
spinor $\hat\epsilon$ is independent of the light cone coordinate $v$,
{\it i.e.}, $\partial_+\hat\epsilon=0$. Thus, we have 
$\hat{\epsilon} = \hat{\epsilon} (u,X^i)$. 

We will now turn our attention to the variations 
$\delta\hat{\psi}_i \ , \ i = 1,..., 8$.
The condition $\delta\hat{\psi}_i=0$ implies 
\begin{eqnarray}  \label{psi-i} 
&\partial_i \hat{\epsilon}
  = \Gamma^- \mathcal{R}  \ \hat{\epsilon} \ , 
\end{eqnarray}
where we have introduced the notation 
\begin{eqnarray} \label{R} 
\mathcal{R} =   \frac{1}{4} f_1(u) \Big(\delta_{i8} \Gamma^7 - \delta_{i7}\Gamma^8\Big)  \sigma_3 + \frac{e^{\hat{\Phi}}}{8} f_2(u) \Gamma^8 \big(i\sigma_2\big)  \Gamma^i  \ . 
 \end{eqnarray}
A quick inspection of the above expression shows that $\Gamma^-$ anticommutes 
with $\mathcal{R}$. Thus \eqref{psi-i} becomes
\begin{eqnarray}  
&\partial_i \hat{\epsilon}
  =  \mathcal{R}   \Gamma^- \hat{\epsilon} \ .
\end{eqnarray}
Since $\Gamma^-  \hat{\epsilon} = 0$, we find $\partial_i \hat{\epsilon}=0$.
Thus, $  \hat{\epsilon}=\chi(u)$ for some $\chi(u)$ such that 
$\Gamma^-  \chi(u) = 0$. This leaves behind us to verify the only remaining 
condition $\delta\hat{\psi}_-=0$.
 %
After some simplification this condition  gives rise to
\begin{eqnarray}  \label{cond} 
 \partial_{u} \chi(u)  - \frac{1}{4} f_1(u) \Gamma^{78}  \sigma_3 \chi(u) + \frac{e^{\hat{\Phi}}}{4} f_2(u)  \Gamma^8 \big(i\sigma_2\big)   \chi(u) = 0 \ . 
\end{eqnarray}
Introducing the matrix
\begin{equation}
{\mathcal{M}}(u) = \frac{1}{4} \left(
f_1(u) \Gamma^{78}  \sigma_3 -  e^{\hat{\Phi}} f_2(u)  \Gamma^8 \big(i\sigma_2\big)\right)
\end{equation}
the above condition can be rewritten as  
\begin{equation}
\partial_u\chi(u) - {\mathcal M}(u) \chi(u) = 0 \ .
\end{equation} 
This equation can be integrated to give rise 
$\chi(u) = e^{\int du {\mathcal M}(u)} \chi_0 .$
Thus the gravitino and dilatino variations are compatible with each 
other for the above choice of $\chi(u)$ along with $\Gamma^-\chi_0=0$.
Since the later condition retains sixteen components of the Killing spinors, 
we find that the pp-wave background \eqref{in X coordinates}-\eqref{Fij}
preserves 16 supercharges. It is interesting to note that there is an 
enhancement of the number of supersymmetries preserved by the pp-wave 
solution originated from the non-Abelian T-dual background. This was the case 
even for the non-Abelian T-dual of the Klebanov-Tseytlin background \cite{Roychowdhury:2021lji}. 
Here the pp-wave solution preserves the supersymmety of the original type $IIB$ supergravity theory on $AdS_3 \times S^3 \times T^4$
before applying T-duality.

\section{Field theory dual}

The field theory dual corresponding to the non-Abelian T-dual background 
\eqref{natdm}-\eqref{natdrr} has been studied in detail in \cite{Lozano:2019ywa}. The construction of the dual theory is based upon the analysis
of quantized brane charges. The T-dual geometry includes a nontrivial 
$S^2$ in the transverse space which interpolates between $\mathbb{R}^3$ 
and $\mathbb{R} \times S^2$ as we move from $\rho\rightarrow 0$ to 
$\rho\rightarrow\infty$. Presence of this nontrivial two cycle allows one 
to construct an intersecting brane configurations involving $D2$ and $D6$ 
branes stretched between $NS5$ branes. The dual gauge theory consists of
a two dimensional $(0,4)$ quiver theory consisting of two infinite 
family of nodes with gauge groups of increasing rank with no flavor
\cite{Lozano:2019ywa}. In support of this construction, the authors 
of \cite{Lozano:2019ywa} computed the central charge of the quiver
theory and shown its agreement with the holographic central charge
computed from the gravity dual.

The pp-wave background obtained upon taking the Penrose limit will 
correspond to a class of operator of the above quiver gauge theory.
To understand it better we will consider the brane charges for the 
pp-wave background \eqref{metric brinkmann}-\eqref{Brinkmann RR}.
The Page charges of various D-branes in type $IIA$ supergravity 
are given by \cite{Whiting} 
\begin{eqnarray}  \label{page charge} 
&\hat{Q}_{\text{page} , D6} =& \frac{1}{2 \kappa_{10}^2 T_{D6}} \int_{C_2}  \hat{F}_2 - \hat{F}_0 \hat{B}_2  \ ,  \nonumber\\ 
&\hat{Q}_{\text{page} , D4} =& \frac{1}{2 \kappa_{10}^2 T_{D4}} \int_{C_4}  \hat{F}_4 - \hat{B}_2 \wedge \hat{F}_2 + \frac{1}{2} \hat{F}_0 \hat{B}_2 \wedge \hat{B}_2 \ ,  \nonumber\\
&\hat{Q}_{\text{page} , D2} =& \frac{1}{2 \kappa_{10}^2 T_{D2}} \int_{C_6}  \hat{F}_6 - \hat{B}_2 \wedge \hat{F}_4 + \frac{1}{2} \hat{B}_2 \wedge \hat{B}_2 \wedge  \hat{F}_2  - \frac{1}{6}  \hat{F}_0  \hat{B}_2  \wedge \hat{B}_2 \wedge \hat{B}_2           \ , \nonumber\\
&\hat{Q}_{\text{page} , D8} =& \frac{1}{2 \kappa_{10}^2 T_{D8}} \int  \hat{F}_0 \ . 
\end{eqnarray}
Similarly, the Maxwell charges are given as \cite{Itsios:2013wd}
\begin{eqnarray}  \label{maxwell charge} 
&\hat{Q}_{\text{Max} , D6} =& \frac{1}{\sqrt{2} \pi^2 } \int_{C_2}  \hat{F}_2   \ ,  \nonumber\\ 
&\hat{Q}_{\text{Max} , D4} =& \frac{1}{\sqrt{2} \pi^2 } \int_{C_4}  \hat{F}_4 \ ,  \nonumber\\
&\hat{Q}_{\text{Max} , D2} =& \frac{1}{\sqrt{2} \pi^2 } \int_{C_6}  \hat{F}_6          \ , \nonumber\\
&\hat{Q}_{\text{Max} , D8} =& \sqrt{2} \int  \hat{F}_0 \ . 
\end{eqnarray}
Here $C_n$ corresponds to appropriate $n$-cycle in the associated geometry. 
In addition, we have the $NS5$ branes with charge
\begin{eqnarray}  \label{NS5} 
\hat{Q}_{NS5} = \frac{1}{4 \pi \alpha^\prime} \int_{C_3}  \hat{H}_3        \ . 
\end{eqnarray}
Subsittuting the values of the background fields for the pp-wave solution
\eqref{metric brinkmann}-\eqref{Brinkmann RR}, we find the non-vanishing
charges 
\begin{eqnarray} 
&\hat{Q}_{\text{page} , D6} = &\hat{Q}_{\text{Max} , D6} = \frac{1}{2 \kappa_{10}^2 T_{D6}} \int_{C_2}  \hat{F}_2   \ ,  \nonumber\\ 
&\hat{Q}_{NS5} =& \frac{1}{4 \pi \alpha^\prime} \int_{C_3}  \hat{H}_3 \ ,
\end{eqnarray}
where the expression of $\hat{H}_3$ and $\hat{F}_2$ are given in \eqref{NS-NS Brinkmann} and \eqref{Brinkmann RR} respectively. This indicates that at the
Penrose limit we are left with only $D6$ and $NS5$ branes. Thus the dual 
gauge theory will be described by an intersecting configuration of 
$D6$ and $NS5$ branes. The BMN operators corresponding to the holographic
dual of the pp-wave geometry will reside in this quiver theory.

\section{Conclusion}

In this paper we have studied pp-wave geometries in non-Abelian T-dual
backgrounds with an $AdS_3$ factor. We focused on the non-Abelian T-dual
of type $IIB$ supergravity theory on $AdS_3\times S^3\times T^4$
background. The resulting T-dual background consists of $AdS_3\times M_3
\times T^4$, where the three dimensional space $M_3$ consists of a 
$S^2$ fibration over a half line. Denoting $\phi$ to be the azimuthal
coordinate on $S^2$ and $\rho$ to be the coordinate parameterizing the
half line, we considered null geodesics carrying non-zero angular 
momentum along the $(\rho,\phi)$ plane. We considered the Penrose 
limit in the vicinity of such null geodesics and showed that the 
resulting geometry gives rise to a pp-wave solution. We solved the 
spinor conditions for this pp-wave background and showed that it 
preserves sixteen supercharges. Finally we commented on the possible
field theory dual for our pp-wave background. It would be interesting
to explore the possibility of exploring pp-wave backgrounds in other
non-Abelian T-dual geometries admitting $AdS_3$ factors. We hope 
to report on this in future.

\vskip .3in
 
\noindent {\bf\large Acknowledgement}

 \vskip .2in

\noindent
The work of SRC is partially supported by grants from the Infosys Foundation to CMI.

\vskip .2in

\appendix 

\section{Einstein's Equations} \label{A1}

Here we will analyse the Einstein's equations for the pp-wave background
\eqref{in X coordinates}-\eqref{Fij}. For type $IIA$ supergravity the 
Einstein's equations are given by
\begin{eqnarray} \label{Einstein eq natd1}
\hat{R}_{\mu\nu} + 2D_{\mu}D_{\nu}\hat{\Phi} = \frac{1}{4} \hat{H}_{\mu\nu}^2 + e^{2\hat{\Phi}} \Bigg[\frac{1}{2} (\hat{F}_2^2)_{\mu\nu} + \frac{1}{12} (\hat{F}_4^2)_{\mu\nu} - \frac{1}{4} g_{\mu\nu} \Big(\hat{F}_{0}^2 + \frac{1}{2} \hat{F}_{2}^2 + \frac{1}{4!}\hat{F}_{4}^2 \Big)\Bigg]  \ .  
\end{eqnarray}
The equation of motion corresponding to the dilaton is 
\begin{eqnarray} \label{R eq natd1}
\hat{R} + 4D^2\hat{\Phi} - 4(\partial\hat{\Phi})^2 - \frac{1}{12}\hat{H}^2=0 \ .
\end{eqnarray}
In the above equations we have used the conventions of \cite{Itsios:2012dc}. In particular, 
we denote 
$\hat{H}_{\mu\nu}^2 = \hat{H}_{\mu\alpha\beta} g^{\alpha\rho}g^{\beta\sigma} \hat H_{\nu\rho\sigma}$ and similar expressions for $(\hat{F}_2^2)_{\mu\nu}$ 
and $(\hat{F}_4^2)_{\mu\nu}$. 

Let us first consider the equation of motion corresponding to the 
dilaton \eqref{R eq natd1}. We will show that each of the terms
in this equation vanish identically. For the pp-wave background 
$\hat R = 0$. To compute
the second term, consider 
\begin{eqnarray} 
&D^2 \hat{\Phi} =& g^{\mu\nu} D_{\mu} D_{\nu} \hat{\Phi} = g^{uv} D_{u} D_{v} \hat{\Phi} + g^{vu} D_{v} D_{u} \hat{\Phi} + g^{vv} D_{v} D_{v} \hat{\Phi} + g^{ij} D_{i} D_{j} \hat{\Phi} \ . 
\end{eqnarray}
Using $\partial_v\hat\Phi = 0 = \partial_i\hat\Phi$ for the partial derivatives
and $\Gamma_{uv}^u = \Gamma_{vv}^u = \Gamma_{ij}^u = 0 $ we find
$D^2\hat\Phi=0$.
Similarly, we can show that $\Big(\partial \hat{\Phi}\Big)^2$ also vanishes identically:
\begin{eqnarray} 
\Big(\partial \hat{\Phi}\Big)^2 =& g^{\mu\nu} \partial_{\mu} \hat{\Phi} \partial_{\nu} \hat{\Phi} = 2 g^{uv} \partial_{u}  \hat{\Phi} \partial_{v} \hat{\Phi} + g^{vv} \partial_{v}  \hat{\Phi}  \partial_{v} \hat{\Phi} + g^{ij} \partial_{i}  \hat{\Phi} \partial_{j} \hat{\Phi} = 0 \ . 
\end{eqnarray}
Finally,  from the expression for $\hat H_3$ \eqref{NS-NS Brinkmann}, it is straightfoward to see that $\hat H_3^2=0$. 
This shows that the dilaton equation \eqref{R eq natd1} is satisfied trivially.

We will now consider the Einstein's equations \eqref{Einstein eq natd1}. 
For our background $\hat{F}_0 = 0 = \hat{F}_4$ and though $\hat{F}_2$ is 
nonzero, from \eqref{Brinkmann RR} we can see that $ \hat{F}_2^2 = 0$. 
Further, a straightforward calculation shows that the only the 
$uu$-components of $ \hat{H}_{\mu\nu}^2 \ ,    (\hat{F}_2)^2 _{\mu\nu}$ together with $D_{u} D_{u} \hat{\Phi}$ are non-vanishing. Similar 
result holds for the Ricci tensor in in Brinkmann coordinates \cite{blau.m}, 
i.e., $\hat R_{uu}$ is the only nonvanishing component of the Ricci tensor. 
Thus, for the pp-wave background \eqref{in X coordinates}-\eqref{Fij}
the Einstein's equation \eqref{Einstein eq natd1} reduces to 
\begin{eqnarray} \label{Einstein uu1}
\hat{R}_{uu} + 2D_{u}D_{u}\hat{\Phi} = \frac{1}{4} \hat{H}_{uu}^2 + \frac{1}{2} e^{2\hat{\Phi}} (\hat{F}_2^2)_{uu}   \ .
\end{eqnarray}
Each of the terms in the above equation can be evaluated in a 
straightforward manner. We find 
\begin{eqnarray} \label{Ruu}
& \hat{R}_{uu} =&  \Bigg[2+ \frac{(4+\rho^2)^2}{\rho^4} J^2 - F_x - F_w  \Bigg] \ , 
 \cr
&e^{-2\hat{\Phi}}=&  \frac{1}{4 \tilde{g}_{s}^2} \big(4 + \rho^2\big)  \ , \nonumber\\
&D_{u} D_{u} \hat{\Phi}=& 4 \ \frac{4\big(\rho^2-4\big) - \big(4 + \rho^2\big) J^2}{\big(4 + \rho^2\big)^2 }   \ , \nonumber\\
&\hat{H}_{uu}^2=& 2 \ \frac{4\rho^2 - \big(4 + \rho^2\big) J^2}{ \big(4 + \rho^2\big)^3 }  \ \Bigg[\big(12 + \rho^2\big)c_3 + \rho \big(4 + \rho^2\big) c_3^{\prime} \Bigg]^2   \Bigg[\frac{c_6^2}{4} + \frac{\rho^2}{4+\rho^2} c_3^2\Bigg]^{-1}      \ ,    \nonumber\\
&(\hat{F}_2)^2 _{uu}=& \frac{J^2}{4\tilde{g}_s^2}   \big(4 + \rho^2\big)                  \ .               \nonumber\\
 \end{eqnarray}
Clearly, substituting the above expressions in \eqref{Einstein eq natd1}
we can see that unlike the remaining equations, it is not satisfied 
identically. However, notice that this equation does involve the only
undetermined coefficient $c_3$ in the expansion \eqref{expand}, and its 
derivative. Thus, it gives rise to a first order differential equation 
involving $c_3$. This equation can be integrated to write an exact 
expression for $c_3$ in closed form.

\end{document}